\documentclass[letterpaper,twocolumn,10pt]{article}
\usepackage{usenix2019_v3}
\usepackage{graphics}
\usepackage{subcaption}
\usepackage{caption}
\usepackage{booktabs}
\usepackage{tikzducks}
\usepackage{fontawesome5}
\usepackage{multirow}

\usepackage{tikz}
\usepackage{amsmath}

\usepackage{filecontents}
\usepackage{xcolor,colortbl}
\usepackage{array}
\usepackage{arydshln}

\newcommand*{\belowrulesepcolor}[1]{%
  \noalign{%
    \kern-\belowrulesep 
    \begingroup 
      \color{#1}%
      \hrule height\belowrulesep 
    \endgroup 
  }%
} 
\newcommand*{\aboverulesepcolor}[1]{%
  \noalign{%
    \begingroup 
      \color{#1}%
      \hrule height\aboverulesep 
    \endgroup 
    \kern-\aboverulesep 
  }%
} 

\newcolumntype{C}[1]{>{\centering\arraybackslash}p{#1}}

\newcommand{\change}[1]{{\color{black}#1}}
\newcommand{\finalchange}[1]{{\color{black}#1}}

\begin{document}

\date{}

\title{\Large ``If sighted people know, I should be able to know:'' Privacy Perceptions of Bystanders with Visual Impairments around Camera-based Technology}

 \author{
 {\rm Yuhang Zhao$^1$,
 Yaxing Yao$^2$,
 Jiaru Fu$^1$,
 Nihan Zhou$^1$
 
 }\\
$^1$University of Wisconsin--Madison,
$^2$University of Maryland, Baltimore County
} 

\maketitle

\begin{abstract}
Camera-based technology can be privacy-invasive, especially for bystanders who can be captured by the cameras but do not have direct control or access to the devices. The privacy threats become even more significant to bystanders with visual impairments (BVI) since they cannot visually discover the use of cameras nearby and effectively avoid being captured. While some prior research has studied visually impaired people's privacy concerns as direct users of camera-based assistive technologies, no research has explored their unique privacy perceptions and needs as bystanders. We conducted an in-depth interview study with 16 visually impaired participants to understand BVI's privacy concerns, expectations, and needs in different camera usage scenarios. A preliminary survey with \change{90 visually impaired respondents and 96 sighted controls was conducted to compare BVI and sighted bystanders' general attitudes towards cameras} and elicit camera usage scenarios for the interview study. Our research revealed BVI's unique privacy challenges and perceptions around cameras, highlighting their needs for privacy awareness and protection. We summarized design considerations for future privacy-enhancing technologies to fulfill BVI's privacy needs.


\end{abstract}

\section{Introduction}

Camera-based technology is becoming increasingly ubiquitous in our daily life. Besides standalone cameras that can be used to capture photos and videos (e.g., life logging cameras and security cameras), smart devices, such as smartphones and smartglasses, also incorporate cameras as input sensors to support different applications. However, while supporting powerful functionalities, camera-based technology can pose significant privacy threats to people, especially to bystanders. 

We define \textit{bystanders} as people who do not have direct control or access to the camera, but can be affected (e.g., captured or recognized) by the sensor or device. In contrast to the \textit{direct users} of cameras, bystanders do not have control over the camera usage and the data it collects and shares; they may not even notice the existence of the camera \cite{bernd2020bystanders}. A bystander’s privacy can thus be easily invaded when someone takes photos that intentionally or unintentionally involve them. The risk is elevated if the photos or videos are shared online. Prior research has explored bystanders' privacy perceptions around different types of camera-based technologies, such as smartglasses \cite{denning2014situ}, logging cameras \cite{hoyle2014privacy, nguyen2009encountering}, and smart home devices \cite{bernd2020bystanders,yao2019privacy}, revealing bystanders' privacy concerns of being captured and their difficulty with distinguishing the camera status \cite{tangipri}. 

Compared to sighted bystanders, bystanders with visual impairments (BVI) face more severe privacy risks around cameras. In many situations, sighted bystanders are capable of noticing the existing cameras, estimating whether they are involved by the camera, and physically avoiding being captured (e.g., walking outside of the estimated area covered by the camera). However, all of these are difficult for BVI due to their vision loss. BVI thus face unique privacy risks and challenges around cameras, leading to different privacy concerns and expectations from sighted bystanders. 

Recently, researchers have realized the importance of the privacy of people with visual impairments \cite{hayes2019cooperative}. With the increasing adoption of camera-based assistive technologies (e.g., Microsoft SeeingAI, Aira), several studies have investigated the privacy concerns of visually impaired people as direct users of such technologies \cite{akter2020privacy, uncomfortables}. Gurari et al. \cite{gurari2019vizwiz} have also constructed a Vizwiz-priv dataset to recognize the private visual information in images to protect the privacy of visually impaired users when they use camera-based technologies to retrieve information from the surrounding environment. However, such prior research for visually impaired people only focuses on their privacy experiences as direct users of cameras. 

Unlike being direct users, visually impaired people can be even more vulnerable when they are bystanders since they do not have control over the device and even cannot visually detect or estimate nearby camera usage. Thus, it is important to ask, how do BVI perceive their privacy when they are around camera-enabled technology? Will they show more tolerance since they understand the importance of cameras as assistive technologies, or will they be more cautious and sensitive due to their vision loss? What unique challenges do they have? What coping strategies do they use to protect their privacy? And what privacy-enhancing technologies do they expect? However, all of these questions remain unanswered. In our research, we aim to fill these gaps by thoroughly exploring BVI's privacy perceptions, expectations, and needs around camera-based technologies. 

We first conducted a scenario-based survey with 90 visually impaired respondents and 96 sighted controls. We designed 12 representative camera usage scenarios by varying three contextual factors, including device types (i.e., smartphone, smartglasses, security camera), camera users (i.e., acquaintances, strangers), and locations (i.e., private, public), and collected BVI's comfort level scores in these scenarios. To further understand BVI's privacy perceptions, we also conducted an in-depth interview study with 16 visually impaired participants from the survey respondents, diving into their privacy concerns, expectations, needs, and willingness to use privacy-enhancing technologies in different scenarios.  

\change{Our survey revealed BVI's challenges in discerning camera usages, as well as their different privacy attitudes from sighted bystanders, especially in smartglass scenarios. Our in-depth interview uncovered BVI's unique privacy perceptions (e.g., low perceived privacy status, strong empathy towards camera users with disabilities) and the complications they faced in a world centered around sighted people (e.g., privacy not respected by sighted people). Lastly, we discuss BVI's privacy needs around cameras and derive design guidelines for accessible privacy-enhancing technologies. }

Our research contributed to the first study on the privacy of bystanders with visual impairments. We revealed BVI's unique privacy perceptions and expectations and built a foundation for the design of accessible and socially acceptable privacy-enhancing mechanisms for BVI.

\section{Related Work}
\subsection{Privacy Concerns of Bystanders around Camera-based Technologies}

Privacy literature has focused primarily on exploring the privacy concerns of the direct users of camera-based technologies \cite{zeng2017end, naeini2017privacy, zheng2018user} and designing privacy-enhancing mechanisms for them \cite{langheinrich2002privacy,yao2019defending}. 


In recent years, researchers have started exploring bystanders’ privacy concerns around different types of camera-enabled technologies, such as lifelogging cameras \cite{hoyle2014privacy, nguyen2009encountering}, drones \cite{flyingdrones, yao2017privacy}, smartglasses \cite{denning2014situ}, and smart home devices \cite{tangipri, bernd2020bystanders, yao2019privacy, marky2020don, lipford2022privacy}. For example, Ahmad et al. \cite{tangipri} found that it was difficult for bystanders to distinguish the `on' and `off' status of the IoT devices even when there were LED lights indicating the camera status. Wang et al. \cite{flyingdrones} revealed that the inconspicuous data collection by the drones and the hidden drone users increased bystanders’ concerns. Moreover, researchers found that bystanders’ privacy expectations varied by context, such as who was collecting the data \cite{lederer2003wants}, whether the space was private or public \cite{flyingdrones}, and the relationship between the users and bystanders~\cite{yao2019privacy}. 

Interestingly, bystanders tend to be more generous in sharing their information when the cameras are used as assistive technologies by people with disabilities. For example, smartglasses were considered more socially acceptable if they were used to support visually impaired users~\cite{ateffect}. However, the information that bystanders were willing to share was limited~\cite{ahmed2018up}. 
\change{Akter et al. \cite{akter2022shared} further revealed the shared ethical concerns between visually impaired camera users and the sighted bystanders, such as the misrepresentation of bystanders by AI technology.}

While prior research explored bystanders’ privacy perceptions around various camera-enabled devices, it mainly focused on bystanders who are sighted. Little attention has been paid to people with visual impairments, who face severe visual challenges and can thus have unique privacy perceptions and needs around cameras. 

\subsection{Privacy Perceptions of People with Visual Impairments as Technology Users}

Because of their vision loss, people with visual impairments face more privacy-related challenges and risks than sighted people.
Hayes et al.~\cite{hayes2019cooperative} found that even though people with visual impairment attempted to seek help from their allies, they were also concerned about the level of privacy, security, and independence they can maintain while sharing personal information with their allies. Moreover, researchers found that visually impaired people had concerns about aural and visual eavesdropping when using screen readers and screen magnifiers \cite{privcbe, azenkot2012passchords}, so that they often used headphones and screen occlusion software to protect their privacy in public. 

Besides traditional aids, visually impaired users also use various advanced camera-based assistive technologies to support their daily activities, such as AI-based technologies (e.g., OrCam \cite{OrCam}, SeeingAI \cite{SeeingAI})
and agent-based technologies (e.g., Be my eyes \cite{beMyEyes}, Aira \cite{Aira}).
Due to their ability to capture users' data, these technologies may bring more privacy issues to visually impaired people. For example, visually impaired users may be concerned about accidentally disclosing their private information to the wrong people when using such technologies \cite{akter2020privacy, uncomfortables} or sharing their information with unknown remote agents~\cite{uncomfortables}.

However, compared to their own privacy, visually impaired users worried more about the bystanders' privacy when using camera-based assistive technologies~\cite{uncomfortables}. For example, with bystanders’ privacy in mind, visually impaired users felt uncomfortable getting some apparent visual information, such as bystanders’ gender and appearance, even though such information was visible by sighted people \cite{akter2020privacy}. Zhao et al. explored visually impaired people's experiences with a face recognition application and also found that they preferred recognizing only their friends to protect others’ privacy \cite{zhao2018face}.



It should be noted that people with visual impairments also face privacy concerns and challenges when they are bystanders of sensing technologies. This, unfortunately, has been overlooked in prior research. We thus fill this gap by investigating the experiences and challenges of people with visual impairments as technology bystanders, exploring how the usage of camera-based technology would affect their privacy perceptions in different scenarios.

\subsection{Privacy-Enhancing Mechanisms to Protect Bystanders from Cameras}

Different privacy-enhancing mechanisms have been designed to protect bystanders from camera-based technologies. Many technologies focused on the post-capturing stage, using image obfuscation techniques (e.g., face blurring or masking) to conceal the private information captured by the camera \cite{frome2009large, hasan2018viewer, wang2017scalable, byspri, li2017effectiveness}. For instance, Hasan et al. ~\cite{obscuringob} categorized ``obfuscation'' into five methods including blurring, pixelating, edge, masking, and silhouette, and compared the effectiveness of each method in obscuring objects in both foregrounds and background. Although such methods can, to some extend, prevent bystanders’ data from spreading across the Internet, they were executed from camera users' end and failed to provide proactive information and controls to bystanders to protect their privacy in the first place. 

Researchers have also designed technologies to increase bystanders’ awareness of surrounding sensors via different visual feedback, such as LED lights \cite{song2020m, byspri, thakkar2022would}, a full-screen flashing camera icon on a computer screen \cite{portnoff2015somebody}, a display showing what the camera is capturing \cite{bellotti1993design}, and visualizing device locations on a map through device registration~\cite{winkler2012user, feng2021design}. Limited research involved audio feedback, such as a beeping sound~\cite{song2020m} or shutter sound~\cite{shutterSound} that indicated camera usage. 

However, most prior privacy-enhancing mechanisms only focused on providing visual indicators for sighted bystanders and overlooked the needs of BVI. To our knowledge, the only privacy-enhancing mechanism designed specifically for visually impaired people was VizWiz-Priv \cite{gurari2019vizwiz}, a dataset that can be used to recognize whether private information was included in a photo sent by visually impaired people, thus providing corresponding notification and protection.
However, it focused on visually impaired people as camera users, not bystanders. Unlike prior research, we aim to focus on BVI and distill design insights to inspire accessible, effective, and socially acceptable privacy-enhancing mechanisms for them.

\section{Study I: A Preliminary Survey}
Given the contextual nature of privacy~\cite{nissenbaum2004privacy, nissenbaum2020privacy}, we conducted a scenario-based survey with \change{90 visually impaired} respondents and \change{96 sighted controls} as our preliminary exploration. \change{Our survey provided quantitative evidence on the major differences between BVI and sighted bystanders regarding their camera detection ability and privacy attitudes in various camera usage scenarios.}
Our research involved minimal risks and received IRB approval to conduct human subjects studies. 

\subsection{Method}
\paragraph{Recruitment \& Eligibility.}
We recruited \change{visually impaired} participants by reaching out to different blind and low vision organizations, such as the National Federation of the Blind (NFB) and Washington Talking Book \& Braille Library (WTBBL), who helped spread out our survey to their members. 
A person was eligible to participate in our study if she or he had visual impairments and was 18 years or older. \change{We compensated the visually impaired participants via a lottery. Respondents had a chance to win a \$20 gift card with a 12\% winning rate. Compared to a fixed-rated compensation, prior study has shown that most participants preferred a lottery and were more motivated to participate with this method \cite{vashistha2015increasing}.} 

\change{We recruited the sighted controls via Prolific \cite{Prolific}, a service that helped researchers recruit participants for their online studies. Since Prolific only supported fixed-rated compensation, we paid each participant \$1.6 for a 9-minute survey. Participants were eligible if they did not have visual impairments and were 18 years or older.} 

\change{We conducted a Cohen's statistical power analysis \cite{cohen2013statistical} to determine the minimum sample size for the study. The results indicated that to achieve 80\% power ($power=.80$) for detecting a medium effect ($f_2=.25$), at a significance criterion of $\alpha=.05$, the minimum required sample size was $N=78$ for general linear models. Thus, the obtained sample sizes of 90 for BVI and 96 for sighted controls were adequate.}


\paragraph{Scenario Selection.}
In our survey, we identified 12 camera usage scenarios to understand BVI's attitudes in different context. The scenarios were constructed based on three contextual factors that could impact people's privacy perceptions: \textit{device type} \cite{tangipri}, \textit{location} \cite{flyingdrones}, and \textit{camera users} \cite{uncomfortables}. We considered different conditions for each factor. 

In terms of device type, we considered three commonly used camera-based devices: smartphones, smartglasses, and security cameras. They covered a wide range of form factors, ranging from handheld to wearable to stationary devices. While smartglasses are less pervasive than smartphones and security cameras, they are emerging technologies that are often used as assistive technologies by people with visual impairments, such as eSight \cite{eSight}, OrCam \cite{OrCam}, and Aira \cite{Aira}. Moreover, these three devices represent different visibility and usage purposes, which may lead to different privacy perceptions. For example, while capturing photos/videos with smartglasses can be much less visible, thus more sensitive than smartphones, using smartglasses as an assistive technology may change people's attitudes towards this device \cite{ateffect}. 

Following prior research on privacy perceptions \cite{flyingdrones, uncomfortables}, we also considered different locations (public vs. private) and BVI's different relationships with the camera users (acquaintances vs. strangers). We then designed 12 camera usage scenarios by traversing all possible combinations of the three contextual factors. All our scenarios reflected real-world stories: they were either commonly encountered scenarios in daily life (e.g., a maintenance person comes to your house and takes photos of facilities) or real stories from news reports (e.g., a stranger wears a pair of smartglasses in a public restroom \cite{CNet}). Appendix Table \ref{tab:scenario} shows all 12 scenarios.

\paragraph{Survey Design.}
Our survey\footnote{The full survey can be found at:~\url{https://drive.google.com/file/d/1SFOzobU2RZMlZFKp2Q5q87uA7whOBjvV/view?usp=sharing}} consisted of three sections: screening, demographics, and scenarios.

In the screening section, we asked whether the participant was visually impaired or sighted, as well as whether they were no younger than 18 years old. We ended the survey automatically if the participant was not eligible for our study.

In the demographic section, we asked about participants' age and gender. \change{If participants had visual impairments, we further asked about their visual ability, including whether they were legally blind, whether they had functional vision, their visual acuity, and their field of view.}  

In the scenario section, we asked questions about the 12 scenarios respectively. For each scenario, we first introduced the scenario and asked the participants to imagine that they were in this scenario. Here's an example of a scenario description: ``Suppose that your friends come to your house to visit you. Some of them use their smartphones to capture photos or videos.'' Participants then gave a 5-point Likert scale score to assess their comfort level in this scenario, where 1 meant extremely uncomfortable and 5 meant extremely comfortable. \change{Similar to prior work~\cite{ackerman1999privacy, cranor2000beyond, lin2014modeling}, we deliberately measured comfort level to infer participants' privacy level and avoided using terms such as ``privacy'' and ``privacy concerns'' to reduce potential bias.}
We then asked about participants' ability to discern camera usage in this scenario with four options, i.e., ``I can always/ sometimes/ seldomly/ never detect whether people are capturing me.'' We repeated the same questions for all scenarios. The order of the scenarios was randomized for each participant to remove the order effect.

We implemented the survey in Qualtrics, an experiment management platform. We also attached the consent form before the survey questions. Participants gave us their consent by continuing with the survey after reading the consent form. Each survey took roughly 5 to 10 minutes to finish. \change{The survey was pre-tested with five sighted pilot participants to ensure that the questions did not have any ambiguity.}

\paragraph{Data Filtering.}  We received 117 responses \change{from visually impaired people} over the course of four weeks (Feb 26, 2021 - March 25, 2021) and \change{120 responses from sighted people over the course of one week (May 9, 2022 - May 15, 2022)}. \change{Based on our prior experience, both the non-profit organizations for visually impaired people and Prolific provide high quality and reliable data entries. As such, we did not add attention check questions.} We filtered the responses by discarding any incomplete or low quality responses (e.g., a visually impaired participant selected 'I can always detect whether people are capturing me' in all scenarios) and responses from ineligible participants. \change{We ended up with 90 responses from visually impaired participants and 96 from sighted participants}.




\paragraph{Data Analysis.}
\change{We detail our analysis methods on evaluating the impact of visual conditions and different contextual factors (i.e., device type, camera user, and location) on participants' comfort level. We defined one measure---ComfortLevel (1 to 5). We had one between-subject factor---\textit{VisualCondition} (two levels: \textit{VI}, \textit{Sighted}), and three within-subject factors---\textit{Device} (three levels: \textit{Smartphone}, \textit{Smartglasses}, \textit{SecurityCamera}), \textit{CameraUser} (two levels: \textit{Acquaintance}, \textit{Stranger}), and \textit{Location} (two levels: \textit{Private}, \textit{Public}).} 
To validate our scenario randomization in the survey, we defined another within-subject factor \textit{Order} to represent the order of each scenario presented to the participant. We also had a random factor \textit{Subject} that represented each participant. 



Since ComfortLevel is an ordinal measure, we fitted our data with a cumulative link mixed model (CLMM) \cite{christensen2018cumulative}. \finalchange{Besides above factors, we also included the interactions that involved VisualCondition in our model since we expected that participants' visual condition may interact with other factors.  Appendix Table \ref{fig:clmm_table} shows the coefficient table of the CLMM model.  
Based on the CLMM model, we calculated the analysis of deviance (ANODE) table with likelihood ratio tests to investigate the effect of different factors on ComfortLevel \cite{mangiafico2016two, christensen2019tutorial}. Appendix Table \ref{tab:ANODE_bothVision} shows the ANODE table.} We found no significant effect of Order on participants' comfort level scores ($ {\chi}^2(11, N=186)= 8.67, p=.652$), which validated the scenario randomization.

\change{We then focused on the data from visually impaired participants, analyzing the effect of different contextual factors on BVI's comfort level. We had one between-subject factor \textit{VICondition} (two levels: \textit{Blind}, \textit{LowVision}). Within-subject factors, measures, and analysis models were the same to the analysis described above. In addition, we analyzed the data from blind and low vision participants separately with the same analysis method to investigate the effect of different factors on the comfort level of BVI with different visual abilities.} \finalchange{Appendix Table \ref{tab:ANODE_VI}, \ref{tab:ANODE_LV}, \ref{tab:ANODE_B} show the ANODE tables respectively.} 

\change{If a factor was found to have a significant effect on the measure based on the ANODE table, we conducted \textit{post hoc} multiple comparisons via Estimated Marginal Means to investigate the relationship between different levels of that factor \cite{mangiafico2016two}. The $p$ value was corrected by Bonferroni Correction.}


\begin{figure*}
    \centering
    \includegraphics[scale=0.135]{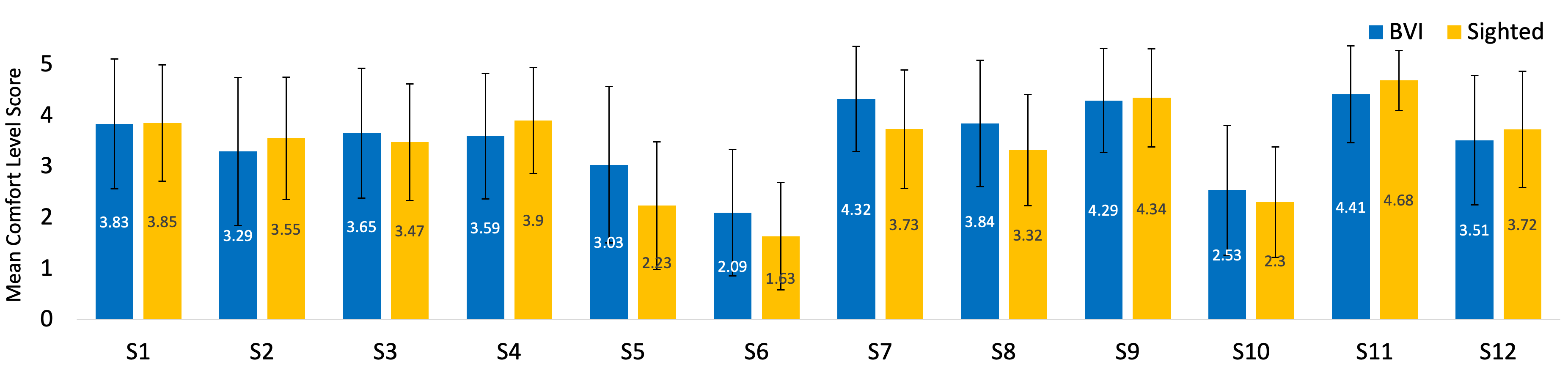}
    \caption{\change{The mean comfort level scores of BVI (blue bars) and sighted participants (yellow bars) across all 12 scenarios.}}
    \label{fig:bar_chart}
\end{figure*}

\subsection{Results}
\label{sec:StudyI}
We received 90 valid responses from visually impaired participants (30 male, 58 female, 1 agender, 1 non-disclosure) whose ages ranged from 19 to 74 ($mean=41.93, SD=13.11$). Among the visually impaired participants, 50 were completely blind, and 40 had low vision. \change{We also received 96 valid responses from sighted participants (30 male, 64 female, 1 queer, 1 non-disclosure) whose ages ranged from 18 to 70 ($mean=34.49, SD=12.51$). We report our results on the differences between BVI and sighted controls across all 12 scenarios, as well as the effect of different contextual factors on blind and low vision participants respectively.}

\textbf{\textit{Ability to detect camera usage.}} \change{We first report participants' camera detection ability. Appendix Figure \ref{fig:camDetect} shows the distribution of participant's camera detection ability in different scenarios. We found that most blind participants can seldomly or never detect cameras in all 12 scenarios (70\%-90\% for each scenario) with ``never'' selected the most in all scenarios (40\%-70\% for each scenario). Compared to blind participants, low vision participants were more capable of detecting cameras with the majority votes felling in ``sometimes'' or ``seldomly'' and only 20\%-30\% participants selecting ``never.'' In contrast, most sighted people (70\%-90\%) indicated that they can at least sometimes detect camera usages in all non-smartglass scenarios.}

\change{Interestingly, we found that the ability to detect cameras was associated with camera types for participants with functional vision (both sighted and low vision). For the sighted controls, most participants reported being able to sometimes detect cameras when the cameras were smartphones (Scenario S1-S4) or security cameras (Scenario S9-S12), but more participants selected ``seldomly'' for smartglass scenarios (S6-S8). Meanwhile, smartphones were easier to detect for low vision participants than smartglasses and security cameras---most participants selected ``sometimes'' in the smartphone scenarios (S1-S3, 50\%-60\%), while the majority votes fell in ``seldomly'' for the rest of the scenarios. We elaborate the reason in Study II.} 


\textbf{\textit{Different privacy attitudes between BVI and sighted bystanders.}} \change{We compared the comfort level between BVI and sighted people in different scenarios. An ANODE analysis based on CLMM model indicated no significant difference between BVI and sighted bystanders ($ {\chi}^2(1, N=186)= 2.13, p=.115 > .05$). However, we found a significant effect of the interaction between VisualCondition and Device on participants' comfort level ($ {\chi}^2(2, N=186)= 35.76, p< .001$). Via a \textit{post hoc} test with correction, we found that sighted participants rated their comfort level significantly lower than BVI in smartglass scenarios ($estimate=-0.898, p<.0001$), indicating that sighted bystanders were more sensitive about their privacy around smartglasses than BVI. Figure \ref{fig:bar_chart} shows the mean comfort level scores of BVI and sighted bystanders in each scenario, where BVI's mean scores were higher than sighted bystanders in all smartglass scenarios (S5-S8).} 

\textbf{\textit{Effects of different factors on the attitudes of blind and low vision bystanders.}} 
\change{We explored the effect of the three contextual factors (device type, camera user, location) on BVI's privacy attitudes. Considering that blind people's privacy perception may differ from low vision people due to their visual ability differences, we analyzed the two groups of participants separately. We first compared the comfort level scores of the two groups with a CLMM model and found no significant difference between blind and low vision participants ($ {\chi}^2(1, N=90)= 0.003, p=.955$). However, by analyzing each group respectively, we found some similarities and differences between the two groups.}  

\change{In terms of similarity, we found that both blind ($ {\chi}^2(1, N=50)= 22.67, p<.001 $) and low vision ($ {\chi}^2(1, N=40)= 9.08, p=.0026 $) participants rated their comfort level significantly lower in a private place than in a public place. Moreover, there was no significant effect of camera users on neither blind ($ {\chi}^2(1, N=50)= 0.47, p=.491 $) nor low vision people's ($ {\chi}^2(1, N=40)= 1.81, p=.178 $) comfort level scores.} 

\change{The differences between low vision and blind participants happened at their attitudes towards camera types. While there was a significant effect of Device on the comfort level of both blind ($ {\chi}^2(2, N=50)= 21.85, p<.001 $) and low vision ($ {\chi}^2(2, N=40)= 6.46, p=.039 $) participants, we found that the relationships between devices varied for the two groups. Specifically, \textit{post hoc} comparison tests with correction showed that, blind bystanders felt significantly more comfortable with security cameras than both smartphones ($estimate=0.599, p=0.0049$) and smartglasses ($estimate=0.881, p<.0001$), and there was no significant difference between smartphones and smartglasses ($estimate=-0.282, p=0.399$). However, instead of security camera, smartphone was the device that low vision participants felt most comfortable with, and they felt the least comfortable with smartglasses. Our analysis indicated that low vision participants rated their scores significantly higher for smartphones than smartglasses ($estimate=-0.552,p=0.034$), but there's no significant difference between security camera and the other two types of cameras (smartphone: $estimate=-0.329,p=0.787$; smartglasses: $estimate=0.313,p=0.454$). The results indicated that low vision people were more sensitive about smartglasses but more comfortable with others using smartphones. This phenomenon relates to low vision people's ability of seeing smartphones, which we explain in Study II.}




\textbf{\textit{BVI's attitudes in different scenarios.}} Finally, we investigated BVI's attitudes in different camera usage scenarios. Participants' mean comfort level scores ranged from 2.09 to 4.41, with ``a stranger using smartglasses in a public bathroom (S6)'' being the least comfortable scenario and ``a museum with security cameras (S11)'' being the most comfortable scenario.   

We identified two top sensitive scenarios, where BVI's mean comfort level scores were below three: (1) \textbf{S6}: a stranger uses smartglasses in a public bathroom ($mean=2.09, SD=1.24$), and (2) \textbf{S10}: a colleague sitting beside you sets up a security camera on his/her desk ($mean=2.53, SD=1.27$). In the former scenario, 26 participants felt somewhat uncomfortable and 39 felt extremely uncomfortable. The latter scenario had 34 participants indicating somewhat uncomfortable and 20 indicating extremely uncomfortable. 


Moreover, we found that BVI's opinions diverged in some scenarios. We used the standard deviation of participants' comfort level scores to measure the divergence of their attitudes. The standard deviation ranged from 0.95 to 1.53 across all scenarios. We identified the top two scenarios where participants presented the highest deviation: (1) \textbf{S2}: a maintenance person uses a smartphone to take photos at your home ($SD=1.45$), and (2) \textbf{S5}: a roommate uses smartglasses in the apartment ($SD=1.53$). \change{Interestingly, both scenarios involved a private place, which indicated that BVI may have controversial opinions towards private places.} 

\section{Study II: An In-depth Interview}
Study I highlighted bystanders' different privacy attitudes due to different camera usage scenarios and different visual abilities.  However, it was not clear how and why BVI's privacy perceptions were impacted by different factors. 
We thus conducted an in-depth interview to thoroughly explore BVI's privacy concerns, perceptions, and expectations around camera-based technologies in different scenarios. 


\subsection{Method}
\paragraph{Participants.}
We recruited 16 visually impaired participants (6 males, 9 females, 1 agender) with ages ranging from 27 to 69 ($mean=45.94, SD=10.78$). All participants were legally blind: 9 were completely blind, and 7 had low vision. All participants had prior experiences with cameras and were familiar with the general camera usage and how it could affect bystanders' privacy. \change{Each participant was compensated with \$15 for a one-hour interview.}

Participants were recruited directly from respondents in Study I so that we can deeply explore what caused their attitudes indicated in the survey. To ensure a good coverage of BVI with different privacy attitudes, we grouped all survey respondents into three groups based on their ``privacy sensitivity'' and recruited participants from each group separately. We measured each respondent's privacy sensitivity by calculating their mean comfort level scores across all scenarios. Respondents' mean scores (i.e., $mean\_comfort$) ranged from 1.08 to 5.0 ($mean=3.53, SD=0.74$). We regarded respondents who gave lower mean comfort level scores as BVI who had higher privacy sensitivity since they were more sensitive to their privacy around camera usage. We thus grouped respondents into three groups: high privacy sensitivity group (i.e., $mean\_comfort < 3$), medium privacy sensitivity group (i.e., $3<= mean\_comfort <=4$), and low privacy sensitivity group (i.e., $mean\_comfort > 4$). Based on our grouping schema, we had 16 respondents who were in the high sensitivity group, 23 low sensitivity, and 51 medium. Given that the high and low privacy sensitivity groups had a much smaller number of respondents, we first sent out recruitment emails to all people in these two groups, resulting in 4 high sensitivity participants and 4 low sensitivity participants. We then recruited from the medium group and obtained another eight participants. \change{Table \ref{tab:demographic} shows participants' demographic information and their privacy sensitivity.}



\begin{table}
\footnotesize
  \begin{tabular}{ p{1.2cm} p{1.2cm}p{2.3cm}p{2cm}}
    \toprule
    Pseudonym & Age/Gender & Visual Ability \newline (Self-reported) & Privacy \newline Sensitivity \\
    \midrule
    James & 40/M  & Blind &  3.58 (M)
    \\
   Mia & 52/F  &	Low vision & 4.33 (L)
    \\
  Emma	& 39/F	 & Low vision & 4.83 (L)
    \\
  Noah & 66/M 	& Blind	& 4.75 (L)
    \\
  Oliver &	49/M & Blind & 3.75 (M)
    \\
   Ella	 & 47/F 	& Low vision & 3.50 (M)
    \\
  Leo &	48/M & 	Low vision & 3.67 (M)
    \\
  Grace & 69/F	 & Low vision & 3.17 (M)
    \\
  Lily & 35/F  & Blind  & 1.83 (H)
    \\  
  Zoe &	45/F 	& Blind &3.67 (M)
    \\
  Olivia & 32/F 	&  Low vision & 2.83 (H)
    \\
  Sofia	& 46/F	 &	Blind & 3.25 (M)
    \\
 Henry & 50/M 	& Blind & 3.67 (M)
    \\
 Alice & 38/Agender  & Low vision & 2.92 (H)
    \\
Dylan &	27/M  &	Blind & 4.75 (L)
\\
Lucy & 52/F 	& Blind  & 2.33 (H)\\
  \bottomrule
\end{tabular}
  \caption{Participants' demographics and privacy sensitivity. 
  }
  \label{tab:demographic}
\end{table}

\paragraph{Procedure.}
\label{sec:procedure}
The study consisted of a remote interview session via Zoom, which lasted 60 to 90 minutes. The interview included three phases: general background, scenario-based questions, and technology design. 

For general background, we asked about participants’ demographic information, visual ability, and experiences using camera-based technologies. We also asked about their real-life experiences as bystanders around cameras, such as their abilities and challenges to detect camera usage, and their coping strategies to protect their privacy. 

For the scenario-based questions, we focused on four scenarios from Study I. We selected the two most sensitive scenarios with the lowest comfort level scores: (1) \textbf{S6}: a stranger uses smartglasses in a public bathroom, and (2) \textbf{S10}: a colleague sitting beside you sets up a security camera on his/her desk, as well as the two most controversial scenarios with the highest deviation: (3) \textbf{S2}: a maintenance person uses a smartphone to take photos at your home, and (4) \textbf{S5}: a roommate uses smartglasses in the apartment. 
For each scenario, we first described the scenario to the participants and asked whether they had encountered similar circumstances in real life. We then asked participants to imagine themselves as bystanders in that scenario and discuss their comfort level (5-point Likert Scale) and the rationales behind it. \change{Participants' mean comfort level score for each scenario remains consistent between Study I and Study II (Pearson correlation: $r=0.87$).} Participants also talked about their ability to detect and avoid nearby cameras, their coping strategies, and their willingness of taking actions against camera usage. 


In the technology design session, we first asked participants’ general concerns about camera-based technology, including how different factors (e.g., location, camera user, camera type) affected their concerns and what information they saw as private and did not want people to capture. We then discussed with the participants about the technology they desired to protect their privacy, for example, what device they wanted to use, what information they wanted to receive, and what feedback can best convey such information.


\paragraph{Data Analysis.}
We recorded the study with Zoom and used Zoom's auto transcription service to transcribe the interviews. \change{Participants had the option to turn off their camera, but they did not have to.} A researcher on the team then went over the recordings and corrected the transcription errors. Two researchers conducted thematic analysis \cite{boyatzis1998transforming} on the interview transcripts. \change{We started the analysis with open coding.} Specifically, two researchers independently coded \change{the same, full transcripts of three participants}, and then came together to discuss the codes. \change{Multiple discussions were conducted to reach an agreement and generate a codebook. Two researchers then split the remaining transcripts and coded them using the codebook. When new codes emerged, they were added to the codebook upon researchers' agreement. Since our coding process involved multiple iterations and discussions, intercoder reliability was not necessary to be checked \cite{mcdonald2019reliability}.}
After open coding, the two researchers met again and derived themes from the codes using affinity diagrams and axial coding. \change{When the initial themes were identified, researchers cross-referenced the original data, the codebook, and the themes, to make final adjustments. After multiple iterations, our analysis resulted in 5 themes and 18 sub-themes\footnote{The full interview protocol and codebook can be found at:~\url{https://drive.google.com/file/d/1SFOzobU2RZMlZFKp2Q5q87uA7whOBjvV/view?usp=sharing}}.} 


\subsection{Findings}





\subsubsection{\change{No Agency to Detect and Avoid Cameras}}
\label{sec:NoAgence}
\change{Due to the vision loss, participants barely had any access to the camera information around them. 
All participants emphasized their lack of ability and agency to visually detect the camera usage by others. As Grace said, \textit{"I never know if I'm in the picture or not. It's really frustrating."} 
While some participants adopted coping strategies to infer camera information via different clues, the current strategies were inadequate.} 



\change{\textbf{\textit{Passively Inferring Camera Usage.}}
Unlike sighted people who can actively scan around to visually identify cameras, BVI were passive information receivers who could only infer camera usage via the audio notifications from the camera or nearby people.  An essential audio cue was the camera shutter sounds mentioned by six participants (e.g., Dylan, James). 
However, such audio cues could be hard to hear if the BVI were not close enough (Dylan) or in a noisy environment (Grace, James, Henry), and the source of the audio could not be easily distinguished (e.g., Henry, Alice).  
More importantly, the audio cues were not always available since they could be easily muted by the camera users (e.g., Oliver, Lucy). 
}


\change{Besides camera shutter sounds, eight participants reported relying on the verbal notifications from the camera users or other bystanders (e.g., overhearing others’ conversations, notified by the camera users). 
However, our participants (e.g., Noah, Zoe, Grace) did not expect to always be informed by others. For example, Zoe did not believe that ``people are going to volunteer'' their camera usage, especially strangers.}  

\change{Due to the lack of awareness of camera information and the ineffective coping strategies, some participants even had a hard time forming their privacy opinions. As Alice mentioned, \textit{“If you're not aware of [the camera], you can't form an opinion about it, [since] it doesn't exist to you.”} }
 
\change{\textbf{\textit{Communication as the Only Way to Protect Privacy.}}
The lack of information about camera usage also largely limited BVI's ability and willingness to take action to protect their privacy. All participants indicated their difficulty in physically avoiding the camera since they could not see the camera position and capture direction. As Olivia said, \textit{``Being a blind person, maybe other people would be more comfortable about [camera usage around them] because they might feel like they would notice, but I don't think I would know. So, it's not like I could change things to put myself in a less uncomfortable situation.''} Thus, participants usually had to fully withdraw themselves from the camera usage environment to avoid being captured (e.g., Sofia, Dylan, Ella). }

\change{One possible way to protect their privacy was through communication. By communicating, confronting, and negotiating with the camera users, BVI tried to collect more camera usage information (Dylan, Henry), reach an agreement with the users (Alice, Dylan), and even stop the users from using cameras (e.g., Noah, Zoe).
For example, Alice described how she would deal with the security camera on her coworker's desk (S10), \textit{``The only way to know really is to just ask the co-worker `what is that'...I would definitely talk to them about it and see if we could come to a resolution. Like, I'm fine with you [using your camera] at this time and see if we could come up with an arrangement.''}} 


\change{However, the lack or uncertainty of camera information made our participants unable (Mia) or unwilling (Olivia) to confront or communicate with camera users. \textit{``I don't [want to communicate]. I don't trust my eyes'' (Olivia).}} 

\change{Moreover, many participants (e.g., Lily, Zoe, Mia) acknowledged that communication would not always be effective. For example, it was hard for them to verify whether the camera users actually honored their requests. As Lily said, \textit{``I would [communicate]. But that still doesn't stop them from keeping doing it. Because it's their word against my word. There still has to be some trust involved. As a totally blind person, there's no way for me to verify that [the camera is off].''}} 

\change{Due to their vulnerability in front of sighted camera users, seven participants demonstrated resignation, and some were even ``not bothered'' to take any actions to protect their privacy (e.g., Noah, Oliver, Leo). As Oliver indicated, \textit{“I have absolutely no control whatsoever over what gets done with any images of me, no matter how they've been acquired. People are going to do whatever they're going to do.
”} 
Some participants (Oliver, Alice, Leo) avoided thinking about the possibility of being captured by cameras since it could induce extra anxiety: \textit{``I tend not to think about [cameras] because I think if I thought about it more, it would be a lot more unsettling.
'' (Alice).}}

\textbf{\textit{Blind vs. Low Vision Bystanders.}}  Unlike blind people, low vision bystanders could sometimes leverage visual cues to estimate the camera usage. \change{For example, some low vision participants (e.g., Alice, Olivia, Mia) mentioned being able to detect camera usage and direction by observing the indicator lights. They also found it relatively easy to detect the use of smartphone cameras by observing camera users' motions and gestures.} As Olivia indicated, \textit{``You hold the phone to get something in the frame or to change the orientation of it, like portrait landscape. I think people’s body language and movements are different.''}

However, low vision bystanders still faced difficulties especially when the cameras were too small or disguised to be visible. 
\change{Moreover, five low vision participants found smartglasses much harder to detect than smartphone cameras.} They had a hard time distinguishing smartglasses from normal glasses, especially when the user was walking around. If the smartglasses did not provide any signals (e.g., an indicator light), they would assume they were normal glasses.  
\change{This echoed the result in Study I that smartglass scenarios were particularly difficult for low vision participants, and also explained why low vision bystanders felt most comfortable in smartphone scenarios and least comfortable around smartglasses.}

\subsubsection{\change{Adapted Privacy Perceptions}}
\label{sec:AdaptedPrivacy}
In our interview, we identified some common privacy perceptions shared between visually impaired and sighted bystanders as suggested by the literature: (1) All participants were more open to public cameras (e.g., security cameras set up by a company or residence building) than personal cameras (a security camera owned by individuals) since they felt more targeted and perceived personal cameras being more likely to capture personal or sensitive information~\cite{flyingdrones}; (2) Most participants (except for Olivia) were more sensitive about private or personal places (e.g., bathroom, bedroom) than public places~\cite{massimi2009understanding, singhal2016you}; and (3) Some participants (e.g., Noah, Lily) tended to compromise to social norms if other people agreed with the camera usage~\cite{thakkar2022would}. However, we also found that BVI presented unique privacy perceptions that differed from sighted bystanders. We elaborate on these unique perceptions. 

\change{\textbf{\textit{Low Privacy Expectation.}} Most participants (11 out of 16) showed low privacy expectations even in sensitive scenarios. They assumed that cameras were everywhere, and as long as they were ``doing the right thing,'' they had no concerns about being captured (e.g., Noah, Leo). Four participants (e.g., Noah, James) reported not expecting any privacy in public places. For example, Leo mentioned that he did not need to know anything about surrounding cameras in public since he would always assume that he was on camera: \textit{``In public, you are as much of an object with everyone around you.''} A more extreme example was that Noah did not care about people using smartglasses in the bathroom at all, even capturing his private part, \textit{``Would it bother me? No, I'm not really very vain. I'm not so different in my private areas from anyone else that it's going to bother me.''} }


\change{With low privacy expectations, five participants (e.g., Alice, Mia) reported constantly monitoring or regulating their own appearance and behaviors to ensure no improper information would be captured by the prevalent cameras. Instead of blaming the camera users, Alice felt herself being responsible if her private information was captured, \textit{``That's on me [and] that was my fault for not moving [personal] things.''} }

\change{\textbf{\textit{High Trust on Acquaintances.}} Twelve participants showed high trust in acquaintances. They assumed that acquaintances would ``know their boundaries'' (Mia) and only ``be doing something nice'' (Sofia) with the cameras. For example, although Olivia felt uncomfortable with a roommate using smartglasses in her apartment (S5), she indicated that she would trust her roommate, \textit{``Even though I'd be very uncomfortable, I know the person. So there's like a level of an implied level of trust, maybe.''} Grace also indicated her willingness to accommodate to her acquaintances. As she mentioned, \textit{``It would be uncomfortable at first, for sure. And I think [your roommate] would have to adjust your way of living. And you'd have to really learn to trust that person.
''}}


\change{Besides high trust, five participants (e.g., Henry, Alice, Ella) felt more comfortable communicating privacy concerns and preferences with acquaintances since they were more likely to give an explanation and respect their requests. As Oliver said, \textit{``If [the camera users] are people to whom I have a personal connection, I am confident that I can convince them to comply [to my privacy preferences] or effectively admonish them for non-compliance.''}}


\change{In contrast, some participants (e.g., Olivia, Noah) were more concerned with strangers using cameras. 
Even for the professional maintenance staff, Dylan and Grace would ask them to show the photos they took before leaving to make sure no personal information was captured. }


\subsubsection{Empathy towards Camera Users with Disabilities}
\label{sec:Empathy}
Empathy was a key factor that drove our participants' privacy attitudes. We found that BVI felt more comfortable if the camera user had a disability and used a camera as assistive technology. \change{However, we also noticed that there were bottom lines that some BVI expected camera users to respect regardless of the purpose of the camera usage.}


\change{\textbf{\textit{High Tolerance towards Camera-based Assistive Technology.}} Most participants (12 out of 16) showed more tolerance for camera users with disabilities since they had similar experiences and believed that people with disabilities were using cameras for assistive purposes. Although they might be accidentally captured, they understood the necessity of the camera usage, \textit{``I feel a lot more comfortable when somebody has a visual impairment, because I know in more detail why they're using that camera, because I have to do the same thing. I know they're using this to gather information about their environment'' (Alice).} }

\change{Interestingly, since smartglasses were a typical assistive device for people with visual impairments to recognize the surrounding environment, participants (e.g., Noah, Emma, Dylan) associated smartglasses with assistive technology directly and showed high acceptance towards the use of smartglasses. As Emma explained, \textit{``If I saw somebody wearing smartglasses, I would think of it as a person that had a disability, so I even have considered like individuals who are deaf if they could [use] wearable technology. I don't think that a person who does not have a disability should be wearing wearable [technology].''} 
The strong association between smartglasses and assistive technology explained the result in Study I that BVI gave significantly higher comfort level scores for all smartglass scenarios than sighted bystanders.} 



\change{\textbf{\textit{Tolerance to an Extent.}} In spite of empathizing with people with disabilities, some BVI's tolerance had a limit. Five participants mentioned that even assistive technology should not be used at very sensitive locations. For example, Dylan was comfortable with visually impaired people using cameras to identify genders at the entrance of public bathrooms but not inside the bathrooms, \textit{``[It] would be good that maybe if you see a blind person walking up to the door and, basically, identifying that this restroom is for this person. And then basically turning off those glasses and going in. I don't feel having such technology in the restroom would be appropriate. I feel sighted people would feel uneasy about that. There's no good agreement where we can get to make this work.''} James and Henry also pointed out that some camera-based assistive technology (e.g., Aira) had deployed policies that prohibited the use of their services in multi-person restrooms.}

Ella and Leo further emphasized the necessity to be notified of the camera usage and confirm the assistive purpose in private locations, \textit{``If they're asking directions, then as long as I can hear that, I guess, the more comfortable is it
'' (Leo).} 

\change{The timing of using cameras was another factor that affected BVI's tolerance. Six participants (e.g., Lucy, James) disagreed with a 24/7 use of cameras. Since the BVI had also used these assistive devices, they can better judge when the devices were needed and when not. For example, Lucy and Alice argued that home was a place familiar to visually impaired people, so there was no need to use cameras constantly, \textit{``I'm visually impaired, [I know] there's not a need for them to have [smartglasses] on all the time
'' (Lucy).}} 

\change{\textbf{\textit{No Additional Tolerance.}}} Four participants (Grace, Lily, Henry, Leo) reported that camera users' disability identity would not affect their attitudes towards camera usage. For example, Grace believed that visually impaired people can still use cameras to do anything besides assistive purpose, \emph{``I don't see what difference it makes. I have used my camera on the phone for a different purpose.''} 

\change{Moreover, some participants showed no tolerance for people with disabilities if they could complete the same tasks without camera assistance. As Henry said, \textit{``I don't have a whole lot of pity and sympathy for [people with visual impairments]. I have no vision, and I can figure out where things are in a bathroom, then you should be able to navigate this bathroom too...You don't need a smartglass in a bathroom to navigate it. That's what a cane is for. Use your cane.''}}




\subsubsection{\change{Complications in a Sighted World}}
\label{sec:Complication}
Our data revealed the complications and inequality faced by BVI regarding privacy and camera technology in a society that is dominated by sighted people.

\change{\textbf{\textit{Privacy Inequity Between BVI and Sighted People.}} Eight participants emphasized the information inequality they faced around cameras compared to sighted bystanders. As Olivia indicated, \textit{``I guess I'm pretty uncomfortable, because I know that other people are better at detecting when there are cameras. And so they have a lot more information, and they can better protect themselves. But I can't. I am not having the same level of information as other people maybe.''} Six participants further expressed their desire for having equal access to the camera information as sighted people have, \textit{``If sighted people know that they are being captured, if everyone knows, then I should be able to know'' (Alice).}} 

 \change{Moreover, four participants (Lily, Oliver, Lucy, Grace) pointed out that, as camera users, they also did not have the same level of agency as sighted people to control the exposure of their camera usage. While sighted people can easily mute their cameras, leaving BVI completely no clues of the camera usage (see details in Section \ref{sec:NoAgence}), visually impaired people's smartphone cameras usually generated audio signals since they had to use screen readers. Moreover, many visually impaired people did not wear earphones because they may prevent them from hearing the surrounding environment \cite{sato2019navcog3}. As a result, the audio feedback from screen readers can easily expose their camera usage to the bystanders. As Lucy described, \textit{``[People] usually [are] able to tell [what I'm doing with my phone]. With a screen reader, if there are no headphones plugged in, it will say where the cameras starting or stopping.''} This emphasized the privacy inequity faced by visually impaired people both as bystanders who cannot equally detect camera usage by sighted people, and as direct camera users who cannot equally hide their camera behaviors in front of sighted bystanders. }

\change{\textbf{\textit{Privacy not Respected by Sighted People.}} Three participants reported their experiences where sighted camera users did not respect their privacy. Lily and Alice reported that they could easily attract people's attention and became the target of cameras when they brought their guide dog. They considered this as a violation of her privacy, \textit{``I don't think it's funny that even if it's something nice, like people have apparently taken videos of me as I praised my dog after crossing the street. I don't think it's cool at all. I think it's still a breach of someone's privacy'' (Lily).} 
This highlighted that photographing guide dogs for entertainment could offend BVI and violate their privacy, especially considering the importance of guide dogs in a lot of visually impaired people's lives.}

\change{Additionally, Dylan described his experience of being captured and posted online without his consent at conventions for blind people, \textit{``I used to attend those conventions for blind folks. And we had some sighted people where they take pictures, and they don't basically tell us and they just post some [photos online], and then I figured out from other people by tagging that I'm in there. I don't like that stuff.''} Taking photos at conventions might be common for sighted people since they could easily notice the photographers, however, taking photos of BVI without notifying them could be a severe disrespect and privacy violation even in public conventions. }



\change{\textbf{\textit{Valuing Others' Privacy above One's Own Privacy.}} As opposed to their own privacy, our participants cared more about other bystanders' privacy and feelings. For example, while Zoe did not mind a blind person using smartglasses in a public bathroom, she argued that it was only acceptable when there were no other bystanders except herself. As she explained, \textit{``
If I knew that [the blind user and I] were the only two people in [the bathroom], then I'd feel a bit differently. But if there were 10 or 15 other people, then I wouldn't [be comfortable].''}  }

\change{Besides other bystanders, participants also cared about the experience of the sighted agents who provided assistive services via camera-based devices\footnote{Some assistive technology (e.g., Aira \cite{Aira}, Be My Eyes \cite{beMyEyes}) connects blind users with a sighted agent via a live-streaming camera, so that the agent can help answer blind users' questions and guide them in specific tasks.}. For example, while Noah did not worry about his own privacy at all in a public bathroom (Section \ref{sec:AdaptedPrivacy}), he mentioned that he wouldn't use a camera-based assistive device since it could not only violate other bystanders' privacy but also offend the remote agent, \textit{``I've been the person with the smartglasses. And when I walked into a bathroom, I had the hope, before I thought about it very much, that an agent might be able to guide me to a urinal or a stall, or a sink, without me having to do a lot of feeling around. But it became very obvious to me that that violates a lot of rights of people who have private parts exposed. And it also has the chance of offending the agent who has to look at that.''} } 

\change{\textbf{\textit{Hoping for Sighted People's Understanding.}} Seven participants expressed the importance of camera-based assistive technology for their lives in ``a sighted world that operates visually'' (Ella). Four participants (Noah, Dylan, Emma, Grace) emphasized their right to use those devices to access equal information, \textit{``When I was thinking about using the smartglasses, I wasn't thinking about taking pictures of people. I was wanting to be able to find the sink, without first touching a urinal. 
If I had a computer program that could recognize a stall, a urinal, a baby changing table, and a sink, I might demand my right to wear glasses in that room for the purpose of detecting those things'' (Noah).}}


\change{Participants expressed the strong desire for sighted bystanders' understanding and hoped that visually impaired people using camera-based assistance would become a widely accepted norm, so that they would not be judged by sighted people. As Dylan envisioned, \textit{``[Visually impaired people] always wonder like what other people think about how we use technology out there. I would like one day to see, our technology to basically make sighted people realize that everything that visually impaired do is normal.''} To achieve this goal, Emma stressed the importance to communicate with others and educate people about camera-based assistive technology, \textit{``[If] I was wearing [cameras] and I go into the bathroom and if somebody is feeling awkward [because] they wouldn't be able to tell that I was blind, so then I'm supposed to be like `hey I don't want you to feel weirded out. I happen to be blind and I'm wearing wearable technology.' Information like this should be communicated right.''} }





\subsubsection{Expected Privacy Enhancements}
Participants also discussed their expectations around cameras and their desired privacy-enhancing technology.

\textbf{\textit{Information Transparency and Certainty.}}
Participants (e.g., Lily, Noah) expected a certain level of transparency shared by camera users (e.g., notifications) ahead of time so that they could take certain actions to avoid being captured. 
As Alice said, \textit{``It's sort of a courtesy to let people know that you're filming them in some respects.''}
\change{Moreover, participants emphasized the importance of information certainty since it was hard for them to confirm the information. An extreme example was that James felt more comfortable if his roommate wore smartglasses the whole day since he would be certain about the camera usage and take actions accordingly. }


\textbf{\textit{Consent.}}
Seven participants (e.g., Mia, Olivia) expected camera users to ask for their consent before capturing them. They were distressed when being captured without their acknowledgment, even in a public convention (Section \ref{sec:Complication}). As Mia explained, \textit{``If it's something that I consented to, if we're having a get together, and the person says, `Hey, smile, I'm gonna take your picture,' at least then I know, they're taking the picture. So at that point, I could say yes or no.''} 


\change{\textbf{\textit{Desired Privacy-Enhancing Technology.}} All participant (except for Leo) expressed their needs for privacy-enhancing technologies. As James said, \textit{``With all the cameras that are around, it would be awesome if I could know what was watching me and when I was in the camera. It would at least give me the autonomy to be able to make my own decision.'' }}

\change{Participants suggested different hardware form factors, such as smartphones (Noah, Ella, Olivia, Mia), smartwatch (Alice), wristband (James), and necklace (Grace), and indicated that the device should be small, invisible, or embedded in technology already used by visually impaired people. \textit{``It is probably better that it's tucked into some other piece of technology like my smartphone or something that I will already be carrying, because as a blind person, it's already hard to figure out what am I going to take with me that I might need and what am I going to leave at home'' (Noah)}.}

\change{In terms of feedback, nine participants preferred proactive notifications about nearby camera usage via audio (e.g., audio beacon from the camera, Ella) or vibration (Grace, Alice, James). However, the feedback should be unobtrusive without distracting themselves or calling attention from others (Alice, Olivia, Noah, James). Emma also wanted an application to detect surrounding cameras and ask for her consent, so that she could have more control over her privacy.}

Moreover, some participants (e.g., Noah, Ella) emphasized keeping the privacy-enhancing technology at a low cost, so that it could be affordable to everyone. 
As Noah indicated, \textit{``What would it cost me to secure this privacy? I'm not sure that I would pay very much. I can pay a few dollars for it, but I wouldn't pay hundreds of dollars for it.''}

\change{\textbf{\textit{Camera Regulation Policies.}} Five participants (e.g., Lily, Dylan, Olivia) suggested deploying camera related policies to protect their privacy. For example, Dylan and Olivia believed that permissions to set up security cameras should be regulated by workspace policies. Dylan also suggested that camera regulation should be part of the employee training to ensure the proper use of cameras by co-workers.  
Moreover, Lily believed that all smartphone companies should force the camera shutter sound so that users were not able to turn it off. Similarly, Dylan mentioned that the device companies could regulate what photos could be taken.}
\section{Discussion}


Our research contributed to the first exploration of BVI's privacy perceptions around camera-based technologies. \change{Our survey (Study I) collected a large amount of data and provided quantitative evidence on BVI's different privacy attitudes from sighted bystanders. Our in-depth interviews (Study II) thoroughly explored BVI's privacy perceptions and uncovered the reasons behind their attitudes towards cameras and camera users.} Our research highlighted BVI's challenges in camera detection and avoidance, and shed light on their privacy expectations and needs for accessible privacy-enhancing mechanisms. We discuss the uniqueness of BVI's privacy perceptions around cameras, as well as design considerations for privacy-enhancing technologies for BVI.

\subsection{Unique Privacy Perceptions of BVI}



\change{One important contribution of our work is to investigate BVI's privacy perceptions around cameras, considering their difficulty of seeing cameras and their special bond with camera-based assistive technologies. On the one hand, we identified some shared attitudes and concerns between BVI and sighted bystanders, for example, cameras were less acceptable in private places than in public places \cite{chowdhury2016bystander, denning2014situ, flyingdrones}, and the purpose of camera usage largely determined bystanders' privacy perceptions \cite{hoyle2015sensitive, hoyle2014privacy}. On the other hand, we uncovered BVI's unique privacy perceptions. We compare our findings with prior literature on sighted bystanders to highlight BVI's uniqueness.}

\change{\textbf{\textit{Perceiving Lower Privacy Status.}} BVI showed low privacy expectations in most camera usage scenarios due to the lack of ability to detect and control others' cameras. While prior work showed a similar pattern that sighted people also resigned or obeyed the emerging trends of ubiquitous cameras \cite{marky2020don, lau2018alexa}, we found that BVI expressed particularly low privacy expectations, even in some sensitive places (Section \ref{sec:Complication}). Many BVI expected no privacy at all in public places, and their definition of public places tended to be broader than sighted people. For example, while sighted people mostly defined bathrooms as sensitive and private places \cite{chowdhury2016bystander, denning2014situ}, some BVI (e.g., Noah, James) perceived the ``public bathroom'' as a public place and did not care about people using cameras in the bathroom. Instead, they focused more on regulating their own behaviors and one (Alice) even blamed themselves for potential information exposure.} 

\change{Moreover, BVI valued others' privacy above their own privacy. Prior work showed that, as the users of camera-based assistive technology, visually impaired people cared a lot about the bystanders' privacy \cite{akter2020privacy, uncomfortables}. Our findings echoed and expanded this result by demonstrating that, even as bystanders, visually impaired people still valued other bystanders' privacy more than their own (Section \ref{sec:Complication}). BVI's perceived low privacy status may come from their long-term living environments and the biased societal culture. Many people with disabilities, as a vulnerable population who rely on assistance from others, live in medical or care settings that preclude their access to privacy (e.g., overprotected parents, a care institute that ignored privacy) \cite{adams2015privacy}. The lack of autonomy and needs for others' assistance may push BVI to perceive privacy as an exclusive privilege for people without disabilities, which lead to their lower privacy expectation and sensitivity. However, more studies are needed to further explore the fundamental reasons behind BVI's low privacy expectations. }

\change{\textbf{\textit{Trusting acquaintances.}} Similar to prior insight that trust in camera owners affected bystanders' privacy perceptions \cite{marky2020don, abdi2021privacy}, BVI in our research highlighted their trust in acquaintances and fewer privacy concerns due to the trust. However, unlike prior work that categorized acquaintances into different levels \cite{abdi2021privacy}, most BVI did not seem to distinguish acquaintances with a high granularity in spite of the different types of acquaintances presented in our scenarios (e.g., friends, family, roommates). Instead, they tended to trust acquaintances in general as long as ``they know them'' (Section \ref{sec:AdaptedPrivacy}). This high trust in acquaintances could be related to visually impaired people's unique relationship with their allies. Since they relied on sighted assistance in many daily tasks, such a relationship helped BVI build more trust towards their acquaintances, even in sensitive camera usage scenarios. However, there could still be privacy tensions between visually impaired people and their allies. Hayes et al.'s work showed that, while visually impaired people depended on their allies to protect their privacy and security, they had privacy concerns especially when their allies might have access to their sensitive information~\cite{hayes2019cooperative}. It is thus important to investigate the social relationships between visually impaired people and their allies to further explore how different relationships with acquaintances may affect BVI's privacy perceptions.}

\change{\textbf{\textit{Empathizing with people with disabilities.}} As users who benefit and even rely on camera-based assistive technology,} BVI showed strong empathy and high tolerance towards camera users with disabilities. Although Profita et al. \cite{ateffect} and Ahmed et al. \cite{ahmed2018up} indicated that sighted bystanders also found cameras for assistive purposes more acceptable, their tolerance was limited (e.g., not willing to share information that exceed the ``natural sight'' of sighted people) \cite{ahmed2018up}. In contrast, BVI in our research demonstrated more understanding and empathy towards people with disabilities. They emphasized the necessity of camera-based assistive technology, including in the bathroom scenario where camera-based technology can help them find the stall (e.g., Dylan, Noah, Emma).     

\change{The perception of smartglasses was the best example that highlighted the differences between BVI and sighted bystanders. In prior literature, sighted people expressed fear of being captured or recorded without their knowledge \cite{googleglass, irelandRayban}. However, when talking about smartglass usage, many BVI (not all) immediately associated smartglasses with assistive technology, especially those who used the technology themselves. Study I also showed that visually impaired participants gave significantly higher comfort level scores than sighted people in all smartglass scenarios (Section \ref{sec:StudyI}), confirming the unique bond between visually impaired people and this particular type of technology. However, we acknowledge that, since smartglasses were new technology, not all participants (e.g., Grace) used or were aware of their potential as assistive technology. In this case, smartglasses were seen as ordinary wearable sensors that did not differ from other cameras.}
 
 \change{\textbf{\textit{Comparing low vision and blind bystanders.}} Unlike people who had little to no vision, low vision people could better estimate the nearby camera presence and usage with their remaining vision. Thus, larger handheld cameras (e.g., smartphone) were usually easier for them to detect than small wearable devices (e.g., smartglasses). Study I showed that smartglasses were the type of camera that low vision people felt least comfortable with (Section \ref{sec:StudyI}). However, we did not find other major differences between blind and low vision people regarding privacy perceptions. The only outstanding case was that one low vision participant, Olivia, mentioned being more privacy-sensitive due to the magnification nature of low vision assistive technology, \textit{``I might be more sensitive as a low vision person, if I'm magnifying something, it's easier for people around me to read what I'm reading. And so I feel like I have to be extra protective.''} 
 More studies are needed to explore the differences between low vision and blind people.}

\subsection{Design Implications}
While perceiving lower privacy status, BVI expressed privacy needs. We draw design implications to inspire accessible and socially acceptable privacy-enhancing technologies to better fulfill BVI's needs and promote privacy equity.




\subsubsection{Information Needs}
\change{Although sighted bystanders may also face privacy difficulties around cameras, the vision loss of BVI significantly intensified the problems. Our results in Study I showed that, while most sighted people can at least sometimes detect camera usage nearby, the majority of blind people can never detect surrounding cameras (Appendix Figure \ref{fig:camDetect}). Compared to sighted bystanders, BVI need more basic, thorough, and certain information about nearby cameras to be able to communicate with the camera users and avoid being captured. We discuss BVI's information needs below.}

\change{\textbf{\textit{Presence of a working camera.}} Camera presence is the most basic information that sighted people can easily detect in many scenarios (e.g., someone taking a photo at an event). Camera usage could also be visually inferred based on the users' posture, indicator lights, and camera behaviors (e.g., some cameras rotate when tracking humans), although the indicator lights could be confusing \cite{tangipri}. However, none of this information is accessible to BVI. Thus, the first goal of a privacy-enhancing mechanism is to notify BVI of working cameras to support basic awareness. }

\change{\textbf{\textit{Camera coverage.}} Whether oneself is covered by a camera is the most essential information BVI care about. Sighted bystanders could estimate whether they are captured by a camera based on the camera position, direction, and the common sense of camera's field of view. Prior work suggested beeping or shutter sound to indicate the camera position \cite{song2020m, shutterSound}, however, this is not sufficient for blind people to determine the camera coverage. 
A privacy-enhancing technique may consider directly informing BVI of whether they are covered by a camera instead of providing only the meta information (e.g., position, field of view).}

\change{\textbf{\textit{Camera users.}} The camera users and their relationships with BVI can affect BVI's privacy perceptions. Our research indicated new camera user categorization beyond existing social relationships (e.g., friends, family \cite{abdi2021privacy})---people who share similar experiences (e.g., the camera user and bystander both have disabilities). However, it was always difficult for BVI to visually recognize others' identity information (e.g., a blind person with a white cane). While future technologies should consider conveying such information to BVI, it is also important to think about whether disclosing such information would invade the privacy of the camera users (e.g., an invisible disability could be private information that people want to hide \cite{santuzzi2014invisible}). }

\change{\textbf{\textit{Camera purpose.}} The purpose of using cameras was important to both BVI and sighted bystanders \cite{thakkar2022would, bernd2020bystanders, denning2014situ}. However, this information was not easy to access for any bystanders, and was potentially hard to recognize via technology. Future research should consider how to collect and accessibly convey such information without violating camera users' privacy.} 

\subsubsection{Technology Needs}
We derive design guidelines to inspire accessible and socially acceptable privacy-enhancing technologies for BVI.

\textbf{\textit{Use convenient devices.}} Convenience remains the top consideration for BVI in terms of privacy-enhancing technology design. Our participants consistently expressed their desire for a built-in application on the devices that they commonly use, such as smartphones. With privacy being a secondary need, BVI do not want to carry extra devices only for the purpose of privacy protection. This is extremely preferable to our participants since, in most cases, they already rely on external devices to assist with their lives.



\textbf{\textit{Provide subtle feedback.}} BVI preferred getting sufficient but subtle feedback, so that the technology would not disturb others. For example, compared to loud audio feedback broadcast to the public, haptic feedback, such as vibration, would be more suitable and socially acceptable. When designing accessible privacy-enhancing mechanisms, designers could leverage different haptic modalities to convey different camera information. For example, different vibration patterns could be designed to indicate whether a camera was used, whether the BVI was in the camera, and whether the camera was taking pictures or recording videos. 

\textbf{\textit{Consider cost and privacy tradeoff.}} Cost-effectiveness ratio is another important factor to consider for privacy-enhancing technologies. Our participants emphasized that an extra device costing much money was not a viable method. Since visual impairments often lead to unemployment and loss of income \cite{mcdonnall2019employment}, the monetary cost of technology, especially privacy-enhancing technology that does not serve functional purposes, could become a major factor that prevents people from adopting it. It is thus essential to consider the cost-benefit tradeoff when designing privacy-enhancing mechanisms for BVI. 


\textbf{\textit{Avoid raising new privacy concerns.}} To design an effective privacy-enhancing system, an important question is: how should we retrieve information about surrounding cameras for BVI? Prior privacy-enhancing mechanisms for sighted people have used crowdsourcing~\cite{das2018personalized, feng2021design} and/or network traffic analysis~\cite{huang2020iot} to provide bystanders information about cameras. Our participants also suggested using computer vision to automatically recognize visual information (e.g., indicator lights on a camera) from the surrounding. However, many information retrieval methods may introduce new privacy concerns, especially those that involve camera usage. It is thus important for designers to take into account all stakeholders' privacy and needs when designing privacy-enhancing mechanisms.

\subsection{Limitations}
Our study has limitations. First, \change{the scenario selection may affect the results in Study I. While we strove to balance different factors (i.e., location, device, camera owner) in the 12 scenarios, it was difficult to guarantee a completely equivalent assignment for each factor across the scenarios. For example, although both bathroom and residency building were categorized as ``private,'' they may indicate different levels of privacy. This could potentially reduce the significant effect of a factor. In fact, even the same place, such as a public bathroom, could be perceived as having different privacy levels by different people. To compensate for this limitation, we conducted a follow-up interview to further understand BVI's privacy perceptions in specific scenarios. Future research should consider involving more balanced scenarios with more participants to remove the potential impact brought by scenario selections.}

Second, participants in Study II were recruited from the participant pool in Study I. We adopted this recruitment strategy intentionally to ensure that our participants covered different privacy sensitivity levels. However, bias may be raised since they've already known all the scenarios. Future research should recruit new participants to further validate the results. 



\section*{Acknowledgments}

We thank the anonymous reviewers for their insightful comments, as well as our study participants for their valuable feedback. This work was partially supported by the University of Wisconsin--Madison Office of the Vice Chancellor for Research and Graduate Education with funding from the Wisconsin Alumni Research Foundation, and the START grant provided by the University of Maryland, Baltimore County.

\footnotesize
\bibliographystyle{plain}
\bibliography{usenix}

\appendix

\section{Appendix}

\begin{table}[!t]
\footnotesize
  \begin{tabular}{p{6cm} p{1.5cm}} 
    \toprule
    \textbf{Scenarios} 	& \textbf{Factors}\\
    \midrule
\textit{S1:} Your friends come to your house to visit you. Some of them use their smartphones to capture photos or videos. & \faIcon{mobile-alt} \hspace{0.1cm}  \faUserFriends  \hspace{0.1cm}  \faHouseUser	 \\
\midrule
\belowrulesepcolor{lightgray} 
\rowcolor{lightgray}\textbf{\textit{S2:} Someone from maintenance comes to your house to check the status of your air conditioner. They use their smartphone to capture photos or videos.}
  & \faIcon{mobile-alt} \hspace{0.1cm}  \faUserAstronaut \hspace{0.1cm}  \faHouseUser
\\
\aboverulesepcolor{lightgray}
\midrule
\textit{S3:} You are attending a meeting in your workplace. Some of your colleagues are using their smartphone to capture photos or videos.
& \faIcon{mobile-alt} \hspace{0.1cm}  \faUserFriends \hspace{0.1cm}  \faUmbrellaBeach
\\
\midrule
\textit{S4:} You are visiting a park, some strangers nearby are using their smartphones to capture photos or videos.
  & \faIcon{mobile-alt} \hspace{0.1cm}  \faUserAstronaut \hspace{0.1cm}  \faUmbrellaBeach
\\
\midrule
\belowrulesepcolor{lightgray} 
\rowcolor{lightgray}\textbf{\textit{S5:} You live with a roommate in an apartment. Your roommate always wears a pair of smartglasses with a camera in the apartment.}  & \faGlasses  \hspace{0.1cm}  \faUserFriends \hspace{0.1cm}  \faHouseUser
\\
\aboverulesepcolor{lightgray}
\midrule
\belowrulesepcolor{lightgray} 
\rowcolor{lightgray}\textbf{\textit{S6:} You walk in a public bathroom in a mall. A stranger in the bathroom wears a pair of smart glasses with a camera~\cite{CNet}. } & \faGlasses  \hspace{0.1cm}  \faUserAstronaut \hspace{0.1cm}  \faHouseUser
\\
\aboverulesepcolor{lightgray}
\midrule
\textit{S7:} You are attending the NFB convention. One of your friends nearby wears a pair of smart glasses with a camera.   & \faGlasses  \hspace{0.1cm}  \faUserFriends \hspace{0.1cm}  \faUmbrellaBeach
\\
\midrule
\textit{S8:} You are visiting a museum. A stranger nearby wears a pair of smart glasses with a camera~\cite{sprigg_2020}. & \faGlasses  \hspace{0.1cm}  \faUserAstronaut \hspace{0.1cm}  \faUmbrellaBeach
\\
\midrule
\textit{S9:} You walk into an elevator in your residency building. There is a security camera setup by the building admin.
 & \faVideo \hspace{0.1cm}  \faUserAstronaut \hspace{0.1cm}  \faHouseUser
\\
\midrule
\belowrulesepcolor{lightgray} 
\rowcolor{lightgray}\textbf{\textit{S10:} You are working in an open office. One of your colleagues who sits beside you set up a security camera on his or her desk.}  & \faVideo \hspace{0.1cm}  \faUserFriends \hspace{0.1cm}  \faUmbrellaBeach
\\
\aboverulesepcolor{lightgray}
\midrule
\textit{S11:} You are visiting a museum. There are security cameras setup by the museum in each exhibition section.  & \faVideo \hspace{0.1cm}  \faUserAstronaut \hspace{0.1cm}  \faUmbrellaBeach
\\
\midrule
\textit{S12:} You are living in an apartment in a residency building. You are familiar with your neighbor across the hallway. One day they installed a security camera at their front door.   & \faVideo \hspace{0.1cm}  \faUserFriends \hspace{0.1cm}  \faHouseUser
\\
\bottomrule
\end{tabular}
\caption{Twelve scenarios in Study I. Highlighted rows represent the four scenarios used in the interview in Study II. We use icons to label the contextual factors of each scenario. \faIcon{mobile-alt}: Smartphones; \faGlasses: Smartglasses; \faVideo: Security Camera; \faUserFriends: Acquaintances; \faUserAstronaut: Strangers; \faHouseUser: Private; \faUmbrellaBeach: Public.} 
\label{tab:scenario}
\end{table}

\begin{table}[t]
\centering
\small
\begin{tabular}{lrrrr} 
\toprule
Level & Estimate & Std. Error & z value & Pr (>|z|)\\ 
\specialrule{.08em}{.2em}{0pt}
 \multicolumn{5}{l}{\cellcolor{lightgray}\textbf{Device} (\textit{baseline = smartphone})} \\
smartglasses & 0.454 & 0.057 & 7.941 & ***\\
security camera & -0.667 & 0.058 & -11.593 & ***\\
\specialrule{.04em}{.2em}{0pt}
\multicolumn{5}{l}{\cellcolor{lightgray}\textbf{Camera User} (\textit{baseline = acquaintance})} \\
 stranger & -0.168 & 0.040 & -4.234 & ***\\
\specialrule{.04em}{.2em}{0pt}
\multicolumn{5}{l}{\cellcolor{lightgray}\textbf{Location} (\textit{baseline = private})} \\  
public & -0.297 & 0.040 & -7.414 & ***\\        
\specialrule{.04em}{.2em}{0pt}
\multicolumn{5}{l}{\cellcolor{lightgray}\textbf{Visual Condition} (\textit{baseline = sighted})} \\  
VI & -0.114 & 0.078 & -1.459 & 0.144\\
\specialrule{.04em}{.2em}{0pt}
\multicolumn{5}{l}{\cellcolor{lightgray}\textbf{Order} (\textit{baseline = order\_10})} \\  
order\_1 & -0.088 & 0.129 & -0.683 & 0.494\\
order\_2 & 0.199 & 0.132 & -1.502 & 0.133\\
order\_3 & -0.105 & 0.133 & -0.787 & 0.431\\
order\_4 & -0.122 & 0.134 & -0.913 & 0.361\\
order\_5 & 0.045 & 0.132 & 0.342 & 0.732\\
order\_6 & -0.056 & 0.129 & -0.434 & 0.664\\
order\_7 & -0.118 & 0.127 & -0.930 & 0.352\\
order\_8 & 0.090 & 0.130 & 0.693 & 0.488\\
order\_9 & -0.091 & 0.131 & -0.695 & 0.487\\
order\_11 & 0.129 & 0.132 & 0.979 & 0.328\\
order\_12 & -0.068 & 0.124 & -0.543 & 0.587\\
\specialrule{.04em}{.2em}{0pt}
\multicolumn{5}{l}{\cellcolor{lightgray}\textbf{Device:Visual Condition} (\textit{baseline = smartphone:sighted})} \\  
smartglasses:VI & 0.171 & 0.057 & 3.008 & **\\
security camera:VI & -0.335 & 0.056 & -5.959 & ***\\
\multicolumn{5}{l}{\cellcolor{lightgray}\textbf{Camera user:Visual Condition} (\textit{baseline = acquaintance:sighted})} \\  
stranger:VI & -0.092 & 0.040 & -2.331 & *\\
\multicolumn{5}{l}{\cellcolor{lightgray}\textbf{Location:Visual Condition} (\textit{baseline = private:sighted})} \\
public:VI & 0.021 & 0.040 & 0.529 & 0.597\\
\specialrule{.00em}{.4em}{0pt}
\hdashline 
\specialrule{.00em}{.4em}{0pt}
\multicolumn{5}{c}{Signif. codes:  '***' p<0.001; '**' p<0.01; '*' p<0.05}\\
\bottomrule
\end{tabular}
\caption{Coefficient table of the CLMM model.}
 \label{fig:clmm_table}
\end{table}
\begin{table}[h]
\centering
\small
\begin{tabular}{lrrr} 
\toprule
Factor & LR Chisq & Df & \textit{p}-value\\ 
\toprule
Device & 147.630 & 2 & *** \\  
Camera User & 19.146 & 1 & *** \\ 
Location & 55.345 & 1 & ***\\
Visual Condition & 2.126 & 1 & 0.145\\
Order & 8.670 & 11 & 0.652\\ 
Device:Visual Condition & 35.764 & 2 & ***\\
Camera User:Visual Condition & 5.440 & 1 & * \\
Location: Visual Condition & 0.280 & 1 & 0.597\\
\specialrule{.00em}{.4em}{0pt}
\hdashline 
\specialrule{.00em}{.4em}{0pt}
\multicolumn{4}{c}{Signif. codes:  '***' p<0.001; '**' p<0.01; '*' p<0.05}\\
\bottomrule
\end{tabular}
\caption{ANODE table when comparing the data from sighted and visually impaired participants. }
 \label{tab:ANODE_bothVision}
\end{table}

\begin{table}[h]
\centering
\small
\begin{tabular}{lrrr} 
\toprule
Factor & LR Chisq & Df & \textit{p}-value\\ 
\toprule
Device & 18.544 & 2 & *** \\  
Camera User & 1.757 & 1 & 0.185 \\ 
Location & 31.726 & 1 & ***\\
VI Condition & 0.003 & 1 & 0.955\\
Order & 14.711 & 11 & 0.196\\ 
Device:VI Condition & 9.079 & 2 & *\\
Camera User:VI Condition & 0.228 & 1 & 0.633 \\
Location: VI Condition & 1.067 & 1 & 0.302\\
\specialrule{.00em}{.4em}{0pt}
\hdashline 
\specialrule{.00em}{.4em}{0pt}
\multicolumn{4}{c}{Signif. codes:  '***' p<0.001; '**' p<0.01; '*' p<0.05}\\
\bottomrule
\end{tabular}
\caption{ANODE table for the data of participants with visual impairments (both blind and low vision). VI Condition includes two levels: LowVision vs. Blind.}
 \label{tab:ANODE_VI}
\end{table}

\begin{table}[h]
\centering
\small
\begin{tabular}{lrrr} 
\toprule
Factor & LR Chisq & Df & \textit{p}-value\\ 
\toprule
Device & 6.465 & 2 & * \\  
Camera User & 1.814 & 1 & 0.178 \\ 
Location & 9.077 & 1 & **\\
Order & 8.939 & 11 & 0.628\\ 
\specialrule{.00em}{.4em}{0pt}
\hdashline 
\specialrule{.00em}{.4em}{0pt}
\multicolumn{4}{c}{Signif. codes:  '***' p<0.001; '**' p<0.01; '*' p<0.05}\\
\bottomrule
\end{tabular}
\caption{ANODE table for low vision participants' data.}
 \label{tab:ANODE_LV}
\end{table}

\begin{table}[h]
\centering
\small
\begin{tabular}{lrrr} 
\toprule
Factor & LR Chisq & Df & \textit{p}-value\\ 
\toprule
Device & 21.851 & 2 & *** \\  
Camera User & 0.475 & 1 & 0.491 \\ 
Location & 22.667 & 1 & ***\\
Order & 12.152 & 11 & 0.352\\ 
\specialrule{.00em}{.4em}{0pt}
\hdashline 
\specialrule{.00em}{.4em}{0pt}
\multicolumn{4}{c}{Signif. codes:  '***' p<0.001; '**' p<0.01; '*' p<0.05}\\
\bottomrule
\end{tabular}
\caption{ANODE table for blind participants' data. }
 \label{tab:ANODE_B}
\end{table}


\begin{figure*}[h]
    \centering
    \includegraphics[scale=0.36]{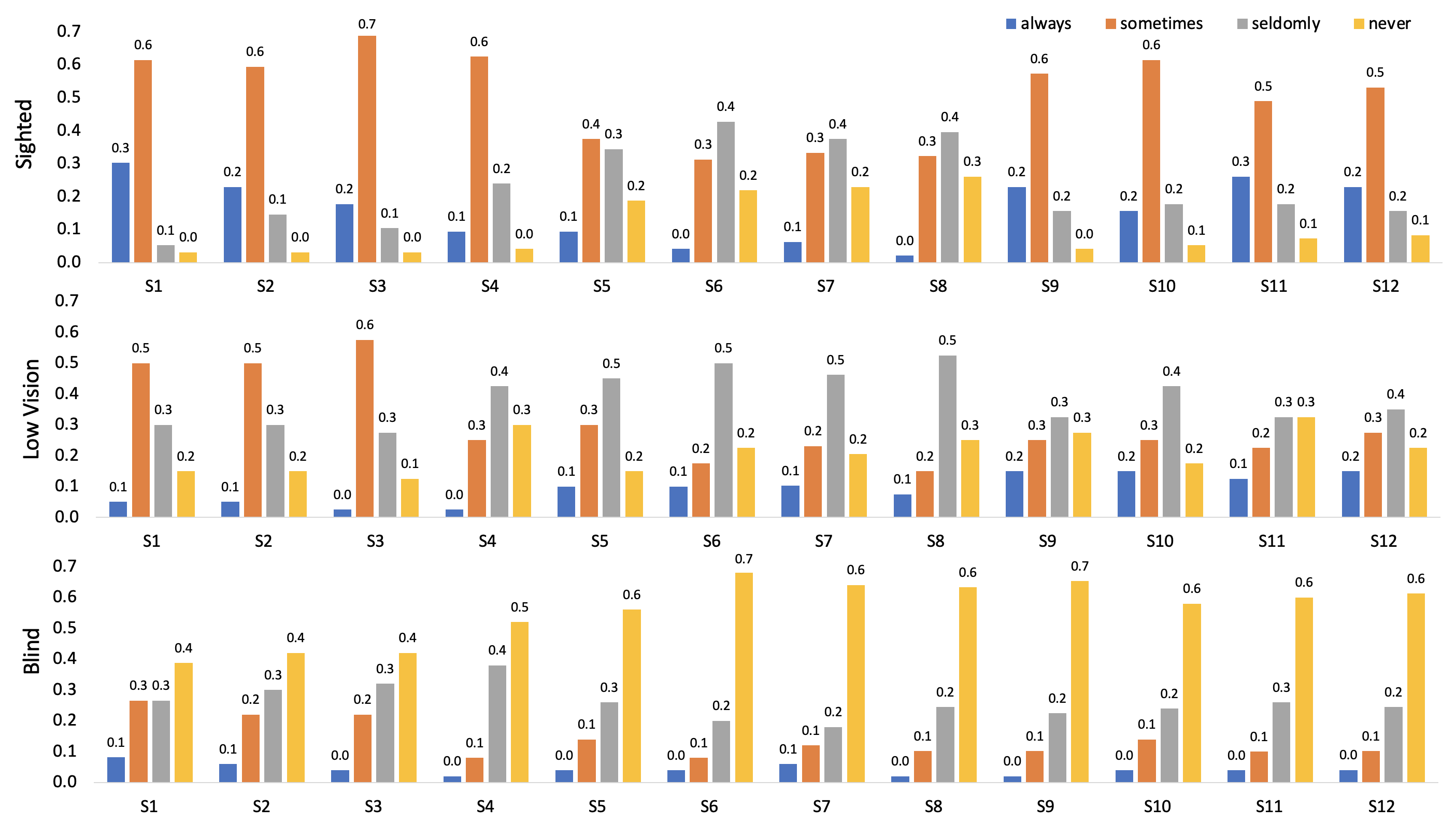}
    \caption{The distribution of camera detection ability of sighted, low vision, and blind participants across the 12 camera usage scenarios. The blue bars represent the percentage of participants who reported to always be able to detect camera usage in the corresponding scenario, oranges bars represent those who select ``sometimes,'' gray bars represent ``seldomly,'' and yellow bars represent ``never.'' The bar charts highlighted that most sighted people could at least sometimes detect camera usage in scenarios where smartphone (S1-S4) and security cameras (S9-S12) were used, however, smartglasses were more difficult for sighted people to detect. Meanwhile, most low vision participants can sometimes detect camera usage in most smartphone scenarios (S1-S3), but they found other scenarios more difficult. In contrast, blind participants faced challenges in general with most of them selecting ``never being able to detect camera usage'' in all scenarios.}
    \label{fig:camDetect}
\end{figure*}

\clearpage


\end{document}